\documentclass[%
 reprint,
superscriptaddress,
nofootinbib,
 amsmath,amssymb,
 aps,
prl
]{revtex4-2}

\usepackage{braket}
\usepackage{graphicx}
\usepackage{dcolumn}
\usepackage{bm}
\newenvironment{nalign}{
    \begin{equation}
    \begin{aligned}
}{
    \end{aligned}
    \end{equation}
    \ignorespacesafterend
}

\usepackage[colorlinks=true, linkcolor=blue, urlcolor=blue,
            citecolor=blue]{hyperref}

\hypersetup{
 colorlinks,linkcolor=blue,citecolor=blue,urlcolor=blue}
 
\begin{document}
\title{Fast Fourier-Chebyshev approach to real-space simulations of the Kubo formula}

\author{Santiago Giménez de Castro}
\affiliation {School of Engineering, Mackenzie Presbyterian University, S\~ao Paulo - 01302-907, Brazil}
\affiliation {MackGraphe – Graphene and Nanomaterials Research Institute, Mackenzie Presbyterian University, S\~ao
Paulo -01302-907, Brazil}

\author{João M. Viana Parente Lopes}
\affiliation{Centro de Física das Universidades do Minho e Porto, LaPMET, Departamento de Física e Astronomia, Faculdade de Ciências, Universidade do Porto, 4169-007 Porto, Portugal}

\author{Aires Ferreira}
\email{aires.ferreira@york.ac.uk}
\affiliation {School of Physics, Engineering and Technology and York Centre for Quantum Technologies, University of York, York YO10 5DD, United Kingdom}

\author{D. A. Bahamon}%
\email{dario.bahamon@mackenzie.br}
\affiliation {School of Engineering, Mackenzie Presbyterian University, S\~ao Paulo - 01302-907, Brazil}
\affiliation {MackGraphe – Graphene and Nanomaterials Research Institute, Mackenzie Presbyterian University, S\~ao
Paulo -01302-907, Brazil}

\begin{abstract}
The Kubo formula is a cornerstone in our understanding of near-equilibrium transport phenomena. While conceptually elegant, the application of Kubo's linear-response theory to interesting problems is hindered by the need for algorithms that are accurate and scalable to large lattice sizes beyond one spatial dimension. Here, we propose a general framework to numerically study large systems, which combines the spectral accuracy of Chebyshev expansions with the efficiency of divide-and-conquer methods. We use the hybrid algorithm to calculate the two-terminal conductance and the bulk conductivity tensor of 2D lattice models with over $10^7$ sites. By efficiently sampling the microscopic information contained in billions of Chebyshev moments, the algorithm is able to accurately resolve the linear-response properties of complex systems in the presence of quenched disorder. Our results lay the groundwork for future studies of transport phenomena in previously inaccessible regimes.
\end{abstract}

\maketitle
Chebyshev polynomials are a cornerstone of spectral approximation theory that have afforded unique numerical toolboxes for investigations of condensed matter \cite{Weisse/2006}. Originally applied to overcome the limitations of power expansion methods in studies of atomic bonding in solids \cite{Wheeler/1972} and to accurately propagate wavepackets in scattering problems \cite{TalEzer/1984}, Chebyshev polynomial expansions of one- and two-particle correlation functions are nowadays routinely used to simulate equilibrium and dynamical properties  
of highly complex systems. This includes, among others, the simulation of localization in percolation models \cite{Schubert/2005}, rare-event effects in disordered semimetals \cite{Pixley_16,Pires/2021}, and excitations in strongly-correlated matter \cite{Holzner/2014,Wolf/2015,Hendry/2021}.

A fundamental trait of spectral approximations is their built-in separation of thermodynamic variables, such as  chemical potential and temperature, from the microscopic information encoded in the\,\,wavefunctions. A prototypical  example is the density of states (DOS) at the Fermi energy \cite{Silver-Roder/1997,Ferreira/2011}. Its Chebyshev\,\,expansion can be cast as $\nu(\varepsilon)= \Phi(\varepsilon)\sum_{n\ge0}\mu_{n}T_{n}(\varepsilon)$, where\,\,\,$T_n({\varepsilon})$ and $\Phi(\varepsilon)=[\pi\sqrt{1-\varepsilon^{2}}]^{-1}$\, are thermodynamic contributions standing for the $n$-th Chebyshev polynomial and weight function in the dimensionless energy domain, respectively. Such an expansion, which is valid for any quantum system with a bounded spectrum, proves most efficient as the bulk of the computational effort is focused on the evaluation of the Chebyshev moments $\mu_{n}=\textrm{Tr}[\,T_n(\hat{h})\,]$ (here, $\hat h$ is a suitably rescaled Hamiltonian with $||\hat h||\le 1$). These moments (hosting the microscopic information) are obtained via a stable sequence of matrix-vector products---circumventing the need for costly exact diagonalization procedures---and stochastic trace estimators may be used to speed up the computation \cite{Iitaka/2003}.

The computational complexity of spectral expansions of one-particle properties calculated for the general class of lattice models with short-range interactions is bounded favourably by $O(DM)$, where $D$ is the Hilbert space dimension and $M$ is the truncation order of the Chebyshev expansion. For non-interacting systems described by the tight-binding approximation, $D$ is the total number of  orbitals and thus the computational effort scales only linearly with the system size. This advantageous $O(DM)$-scaling has been leveraged to enable unprecedented fully non-perturbative studies of one-particle properties in disordered systems \cite{Ferreira/2015, Bouzerar/2020,Varjas/2020,KITE_2020,Lothman/2021,Pires/2022,Joao/2022,Liu/2022, FAN20211}, but open questions remain regarding the possibility to handle two-particle properties (crucial for non-equilibrium studies) in an equally satisfactory fashion. Perhaps the most familiar among these is the electrical conductivity tensor, describing the charge current response to an external electric field \cite{mah00}. The  longitudinal DC response, $\sigma_{aa}:=\lim_{\omega\rightarrow0}\, \sigma_{aa}(\omega)$  (with $a=x,y,z$), is only dependent on the electronic states at the Fermi energy (in the $T\rightarrow 0$  limit). Therefore, it turns out that $\sigma_{aa}$ at the Fermi energy can be computed via the Kubo-Greenwood formula with the same level of efficiency of a DOS calculation, courtesy of an \textit{exact} Green's function Chebyshev polynomial expansion  \cite{Ferreira/2015}. In other words, the number of arithmetic operations is favourably bounded by $O(DM)$. 
However, this strategy hinges upon the separability of the Chebyshev expansions of the retarded and advanced Green's functions in the DC Kubo-Greenwood conductivity due to containing only pure Fermi-surface terms. This is not warranted in more general scenarios, in which the Fermi sea of the electronic system  plays a significant role (i.e., whenever the spectral representation of the response function contains off-Fermi surface contributions arising from transitions between filled and empty states). In fact, while the DC conductivity  of time-reversal ($\mathcal{T}$) invariant systems is a \textit{pure} Fermi-surface property, the breaking of the $\mathcal{T}$-symmetry changes this picture entirely \cite{Baranger/1989}. A well-known example is the quantized anomalous Hall response of Chern insulators \cite{Haldane_04}, which is determined by the Berry curvature of all occupied  bands. Application of the spectral framework yields, in the general case, $\sigma_{ab} =\sum_{n,m} \Lambda^{nm}  \mu_{ab;nm} $. Here, $\Lambda^{nm}$ is a known function of the Fermi energy \cite{Ferreira/2015},  $\mu_{ab;nm}$ is the trace of the special operator string $T_n(\hat h )\hat v_a T_m(\hat h)\hat v_b$ \cite{Weisse/2006}, and $\hat{\textbf{v}}$ is the velocity operator. Evaluation of all the $\mu^{nm}_{ab}$ gives full-spectral access to the conductivity tensor, with any desired energy resolution provided $M$ is large enough \cite{Ferreira/2015}. Unfortunately, such a task traditionally requires $O(DM^2)$ operations \cite{Garcia_PRL/2015}, which severely hampers the classes of  problems that can be  tackled.

In this Letter, we report linear-response transport simulations in graphene and bilayer graphene nanostructures that carry the spectral information of billions of double-Chebyshev expansion moments (i.e., $M^2 = O(10^{10})$). This large-scale computational experiment is carried out with a new approach we present below, whose complexity scales with $D M \log  M$ (in contrast to $O(D M^2)$) and  hence entails a computational effort similar to a standard DOS calculation. This key improvement over the state of the art is shown to enable full-spectral calculations of the conductivity tensor with energy resolutions approaching the mean-level spacing of real systems. 

\textit{Setting the stage.---}We begin by briefly reviewing the main ingredients in the  Chebyshev approach to real-space linear-response transport calculations. The DC conductivity at finite temperature may be cast as  
\begin{equation}
 \sigma_{ab}(\mu,T)=\frac{2e^{2}\hbar}{\pi\Omega}\,\Re\int dE\,f(E,\mu,T)\,\textrm{Tr}[\,\hat{\mathcal{O}}_{ab}(E)\,],
\label{eq1:Kubo-Bastin}
\end{equation}
where $f(E,\mu,T)=1/[1+e^{(E-\mu)/k_B T}]$, $\mu$ is the chemical potential, $\hat{\mathcal{O}}_{ab}(E)=i\hat{v}_{a}[\partial_{E}\,\hat{\mathcal{G}}(E)]\,\hat{v}_{b}\,\Im\,\hat{\mathcal{G}}(E)$ is the conductivity kernel operator, $\mathcal{\hat{G}}(E)=(E-\hat{H}+i0^{+})^{-1}$ is the retarded Green's function, $\widehat{H}$ is the real-space lattice Hamiltonian, $\hat{v}_a$ is the $a$-th component of the velocity operator, and $\Omega$ is the $d$-dimensional volume \cite{Bastin/1971}. $\mathcal{\hat{G}}(E)$ is approximated by a truncated Chebyshev expansion \cite{Ferreira/2015}, smoothed by convolution with a kernel \cite{Weisse/2006}. Hereafter,  we focus the discussion on the diagonal DC conductivity and relegate the more general derivation  to Ref. \cite{supp_material}. For efficiency, the trace in Eq. (\ref{eq1:Kubo-Bastin}) is converted into a stochastic average over random vectors  $\{\ket{r}\}_{r=1,...,R}$ \cite{Iitaka/2003}, according to
\begin{equation}
\sigma_{aa}(\mu,T)\simeq\frac{4e^{2}}{h\Omega}\,\int_{-1}^{1}d\varepsilon\,[-\partial_{\varepsilon}f(\varepsilon,\tilde{\mu},\tilde{T})]\,\langle\langle\tilde{\sigma}_{a}^{r}(\varepsilon)\rangle\rangle_{R}, 
\label{eq2:XXcase}
\end{equation}
where $\braket{\braket{\hspace{1mm}\cdot \hspace{1mm}}}_R$ denotes the statistical average with respect to the $R$ random vectors and $\tilde{\sigma}^r_{a}(\varepsilon)$ is the contribution of a single random vector, formally given by
\begin{equation}
\tilde{\sigma}^r_{a}(\varepsilon)=\sum^{M-1}_{m,n=0} g_{mn}(\varepsilon) \mu^r_{a;mn}, 
\label{eq5:traditionalFS}
\end{equation}
where $\mu^r_{a;mn}= \bra{r}T_m(\hat h) \tilde{v}_a T_n(\hat h)\tilde{v}_a\ket{r}$,  $\tilde{v}_a=\hbar\hat{v}_a/\Delta E$, $g_{nm}(\varepsilon)= g_m  T_m(\varepsilon)\hspace{0.5mm}g_n T_n(\varepsilon)/(1-\varepsilon^2)$, and $\hat h$ and $\varepsilon$ are rescaled versions of $\hat{H}$  and $E$, which map the eigenvalues of $\hat{H}$ and the energy onto the  interval $[-1:1]$ (similar definitions apply to the rescaled temperature and chemical potential variables, $\tilde \mu$ and $\tilde T$). The error in the stochastic trace evaluation is negligible provided that $R D \gg 1$ \cite{Weisse/2006}, which is easily satisfied for the large systems considered here. As a rule of thumb, the relative root-mean-square fluctuation scales as $1/\sqrt{R D}$, and thus can be as low as $1\%$ for just a single random vector realization of a system with $D=10^4$ sites.  Furthermore,  $\Delta E$ is the energy bandwidth of the Hamiltonian and $g_m=2K_m/(\pi(1+\delta_{m,0}))$ encodes the kernel coefficients. In this work, the Jackson kernel \cite{Jackson/1912} is adopted (see Ref. \cite{JacksonKernel_coeffs} for the explicit form of $K_m$). The spectral resolution is determined by the Gaussian width and  satisfies $\delta E \le \pi \Delta E / M$ \cite{Weisse/2006}. 

\textit{The fast Fourier-Chebyshev algorithm (FastCheb)}.--- From the definition of $\mu^r_{a;mn }$ we easily see that the total number of floating-point operations is $O(DM^2)$ for sparse Hamiltonians, validating our earlier claim. In what follows, we explore the powerful fast Fourier transform (FFT) algorithm \cite{CooleyTukey/1969, Frigo/2005} to improve the scalability.
 
First, we split the dimensionless energy interval $-1\le\varepsilon_k\le 1$ into $M$ energy points $\varepsilon_k$, with $k=0, \, ...,\, M-1$ . At each of these points, Eq. (\ref{eq5:traditionalFS}) for the kernel $\tilde{\sigma}^r_{a}(\varepsilon_k)$ can be rearranged as a scalar product between two energy dependent vectors \cite{Ferreira/2015}
\begin{align}
\tilde{\sigma}^r_{a}(\varepsilon_k)= &\frac{1}{ 1-\varepsilon_k^2} \times  \nonumber\\
&\left[\sum^{M-1}_{m=0} g_m  T_m(\varepsilon_k)\bra{a^{L;r}_m}\right] \cdot \left[ \sum^{M-1}_{n=0} g_n T_n(\varepsilon_k)\ket{a^{R;r}_n}\right] \nonumber\\
  =&\frac{1}{1-\varepsilon_k^2}\braket{\phi^{L;r}_{a;k} | \phi^{R;r}_{a;k}}, \label{eq7:dotProd}
\end{align}
which we call left and right vectors. Moreover, we have introduced the auxiliary vectors $|a_{m}^{L;r}\rangle=T_{m}(\hat h)|r\rangle$ and $|a_{n}^{R;r}\rangle=\tilde{v}_{a}T_{n}(\hat h)\tilde{v}_{a}|r\rangle$ for compactness of notation. Next, the energy points are carefully chosen in order to match the nodes of the Chebyshev polynomials of first kind, i.e. $\varepsilon_k=\cos{(\pi (k+1/2)/M)}$. Exploiting the Chebyshev-to-Fourier mapping, $T_n(x)=\cos (n \arccos x)$, the right and left vectors at the $\varepsilon_k$ points become
\begin{equation}
\ket{\phi^{R/L;r}_{a;k}}=\sum^{M-1}_{m=0} \frac{\cos\left(m\pi (k+1/2)/M\right)}{1+\delta_{m,0}}\frac{2K_m}{\pi}\ket{a^{R/L;r}_m}.
\label{eq8:energySpaceVec}
\end{equation}
These are discrete  cosine Fourier transforms of the vector sequences $K_m/\pi \ket{a^{R/L;r}_m}$, which is the main result of this Letter. We call these energy-space vectors. 

Next, we employ a divide-and-conquer strategy to obtain all of the energy-space vectors \textit{simultaneously} in an efficient manner. The key steps are as follows. First, we carry out the vector recursions and construct the matrices $\boldsymbol{a}_{L(R)}$ by lining up all vectors of the left (right) sequences $|a_{n}^{L(R)}\rangle$ along their columns. Then, we run through the matrices $\boldsymbol{a}_{L(R)}$, row by row, performing one-dimensional cosine FFTs. These FFTs yield the $i$-th rows of the energy-space matrices as $\phi_{a;i,k}^{R/L;r}=\mathbf{FFT}^{m\rightarrow k}[K_{m}a_{i,m}^{R/L;r}/\pi]$ ($i=1,...,D$). The partial result $p_{a;i}(\varepsilon_k)$ of the dot product from Eq. (\ref{eq7:dotProd})  is updated every time an energy-space row is obtained, i.e. $p_{a;i}(\varepsilon_k)=p_{a;i-1}(\varepsilon_k)+\phi^{L\dagger}_{a;i,k}  \phi^R_{a;i,k}$. Finally, once all $D$ rows have been visited,  the random vector contribution is obtained as $\tilde{\sigma}^r_{a}(\varepsilon_k)=p_{a;D}(\varepsilon_k)/(\pi^2(1-\varepsilon_k^2)) $. The explicit evaluation of the $M^2$ Chebyshev moments $\{\mu^r_{a;mn}\}$ is bypassed. In its place, $D$ FFTs of length $M$ are performed, yielding a total of $O(DM\log{M})$ operations.

Realizing the full extent of these advantages, however, requires that the $\boldsymbol{a}_{L(R)}$ are all stored in memory. That entails a memory cost $O(2DM)$, which is demanding for large systems \cite{Ferreira/2015}. To overcome this challenge, we employ a partitioning scheme which we discuss in detail in the Supplementary Material \cite{supp_material}.

\begin{figure}
\vspace{3mm}
\begin{center}
\resizebox{8.5cm}{!} {\includegraphics{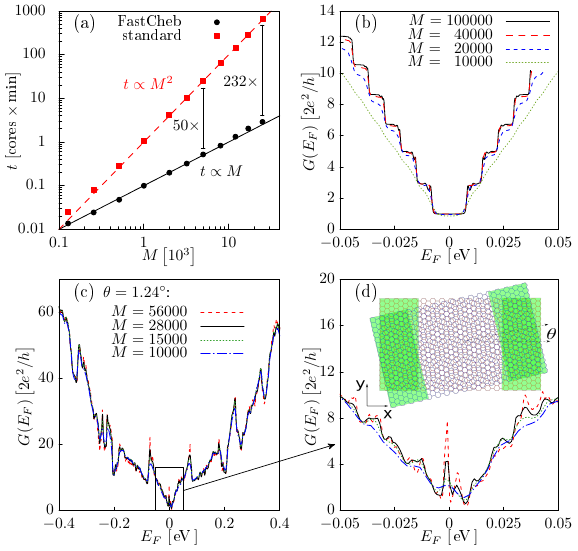}}
\caption{(a) Scaling of CPU time with Chebyshev order $M$ in the \textit{FastCheb} and standard approaches. This benchmark is for a single random vector evaluation of  $\sigma_{xx}$ in a small system with 256$\times$128 orbitals. (b) Fermi-energy dependence of the linear conductance of a large graphene nanoribbon with side lengths $L_x=L_y=100$ nm, to which absorbing contacts of the same dimension are attached. The full TB model contains $10^7$ orbitals. (c) and (d) Panoramic and detailed views of linear conductance curves for a TBG device with a total of $2.5 \times 10^6$ orbitals. Values of $M$ corresponding to each curve in (b)-(d) are indicated on the plots.  See Ref. \cite{supp_material} for more details.}
\label{fig:01}\end{center}  
\end{figure}

\textit{Implementation and benchmark.---}In order to assess its baseline performance, \textit{FastCheb} is implemented within framework of the open-source {\ttfamily KITE} code \cite{KITE_2020}. {\ttfamily KITE} is a high-performance code for spectral simulations of  Green's functions and related quantities in real space \cite{Pires/2022,Joao/2022}, and hence is an ideal testbed for this study. The efficiency of our algorithm can be best appreciated in a direct comparison with the standard recursive method. To this end, we simulate the diagonal conductivity of graphene using a minimal nearest-neighbor tight-binding (TB) model on a honeycomb lattice. The results of this benchmark are summarized in Fig. \ref{fig:01}(a), where the computational effort  is seen to follow  closely the behavior expected from the earlier considerations. In this example, the calculation with $5000$ Chebyshev moments using \textit{FastCheb} is $50$ times faster compared to the standard approach, while for $M=25000$ it has already become $232$ times faster. This $M$-scaling law for the CPU time is robust and representative of a wide class of problems. By the way of two main case studies, we show  below that \textit{FastCheb} has pivotal advantages in linear-response studies of bulk electrical conductivity  and conductance in nanostructures.

\textit{Ballistic conductance and twisting effects}.---We start by simulating a two-terminal quantum-transport device made from a large graphene nanoribbon. The linear conductance at the Fermi energy, $G(E_F)$, is obtained from the $T \rightarrow 0$ limit of Eq. (\ref{eq2:XXcase}) [$\sigma_{aa}(E_{F})\simeq\frac{4e^{2}}{h\Omega}\,\langle\langle\tilde{\sigma}_{a}^{r}(\varepsilon_{F})\rangle\rangle_{R}$] using the  framework for two-terminal devices recently developed in Ref. \cite{Gimenez_2023}. Note that in the zero-temperature limit, this quantity is a pure Fermi surface property and the conductivity kernel directly yields the DC response. The linear conductance  is known to exhibit well-defined quantization steps due the transverse subbands formed in confined nanostructures. The changes in the conductance occur in discrete steps of $e^2/h$ (per spin) each time a new transport channel opens up at the Fermi level \cite{Exp_van_Wees_88,Exp_Wharam_1988}. Figure \ref{fig:01}(b) shows that this behavior is fully captured by the spectral algorithm, with the Chebyshev truncation order playing a central role. Because the plateau width is so small (around 1 meV ),  tens of thousands of iterations are required to accurately reproduce sharp step changes in the conductance. This can be traced back to the slow algebraic convergence generated by step-like discontinuities, one of the hardest singularities to resolve with polynomial expansions \cite{Boyd/2000}. Thanks to the efficiency of \textit{FastCheb}, such fine features can be captured with modest computational effort. 

Next, we demonstrate the power of our algorithm by resolving the linear-response conductance of large twisted bilayer graphene (TBG) devices. We focus on commensurate structures with small twist angle $\theta$, modelled via a Slater-Koster tight-binding scheme \cite{SlaterKoster/1954, Bahamon/2019}. To faithfully capture interlayer interactions in TBG, it is imperative to go beyond a nearest-neighbour approximation \cite{TBLG_TB_Moon_12}. Here, we include full hopping integrals up to a distance of $0.58$ nm (approximately four times the bond length). In practice this entails around $60$ neighbors for each carbon atom, which makes a transport study prohibitively demanding for exact diagonalization. The conductance of a TBG device with $\theta=1.24^\circ$ obtained with \textit{FastCheb} is shown in Figs. \ref{fig:01}(c)-(d). Owing to the full-spectral capability of the algorithm, it is possible to zoom in on small features, such as the conductance peak at the charge neutrality point caused by residual dispersion of the conduction bands. The absence of ballistic conductance steps is due to the channel mixing caused by elastic scattering between the layers, and is thus a direct result of moiré supercell effects \cite{Gimenez_2023}. Based on our extensive tests, we estimate that the  largest TBG simulation (i.e., $M=56000$, corresponding to a spectral resolution of $0.5$ meV, and taking $40$ core hours with \textit{FastCheb}), would require around $2400$ core hours using the standard kernel polynomial approach. 

\begin{figure}
\vspace{3mm}
\begin{center}
\resizebox{8cm}{!} {\includegraphics{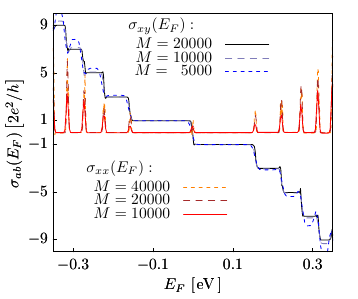}}
\caption{Longitudinal and transverse conductivity as function of the Fermi energy in a disordered graphene system tuned to the quantum Hall regime. The lattice has a total of $10^7$ sites. The parameters are set to $W=0.1t$ (disorder strength), with $t=2.7$eV (nearest-neighbor hopping energy), and magnetic flux $\Phi \simeq 3\times10^{-4} h/e$. Data are averaged over 10 random-vector and 4 disorder realizations. Periodic boundary conditions are employed.}
\label{fig:02}  
      \end{center}  
 \end{figure}


\textit{Kubo-Bastin formulation: Hall conductivity.---}The key motivation for developing \textit{FastCheb} is to extend the range of transport phenomena that can be studied by means of microscopic lattice models. It is well known that many facets of disordered systems and quantum criticality are notoriously challenging to address numerically even  at the level of one-particle properties (such as the inverse localization length   \cite{QH_Kramer/1993,QH_Slevin/2009}). An example is the scaling behavior of integer quantum Hall transitions that remains a long-standing problem, with most recent progress making use of transfer-matrix calculations in quasi-1D  geometry \cite{QH_Obuse/2012,QH_Gruzberg/2017,QH_Puschmann/2019,QH_Slevin/2023}. The possibility to perform  large-scale lattice calculations of the full conductivity tensor in 2D geometry makes  \textit{FastCheb} a promising tool in quantum transport. Here, a start is made towards the application of such a tool to quantum Hall systems. The Kubo-Bastin formalism is employed for this purpose because it provides a unified treatment of all components of the conductivity tensor   \cite{Bastin/1971} and is amenable to spectral expansions \cite{Garcia_PRL/2015, Garcia/2022, KITE_2020,supp_material}.

We choose the integer quantum Hall effect in  graphene to demonstrate the capacity of \textit{FastCheb} to probe topological transport in large 2D systems. The Hall conductivity in graphene obeys an unconventional quantization condition, $\sigma_{xy}^{n}=\mp 2e^{2}/h\,(2n+1)$ with $n\in\mathbb{Z}_{0}^{+}$ (here, the $\pm$ signs hold for electrons/holes), due the  Berry phase of the electron wavefunctions \cite{QH_Graphene_Gusynin/2005,QH_graphene_Zhang/2005,QH_graphene_Novoselov/2005}.  The numerical implementation of the Kubo-Bastin formula revolves around the same concepts as before, however this time 6 energy-space vectors are required (as opposed to  two for pure Fermi surface quantities like the $T=0$ longitudinal conductivity). The perpendicular magnetic field is included in the Hamiltonian through Peierls' phases in the hopping terms, generating a magnetic flux $\Phi$ per unit cell. To emulate the effect of disorder, we supplement the TB Hamiltonian with an uncorrelated on-site potential \cite{KITE_2020}.

The longitudinal ($xx$) and Hall ($xy$) conductivity of a large system with side lengths $L_x=1600$ and $L_y= 3200$ (in units of the lattice spacing) is shown in Fig.\,\ref{fig:02}. Both quantities are seen to follow the expected behavior for the insulating quantum Hall regime of graphene \cite{QH_Graphene_Gusynin/2005}, thus validating the robustness of the new approach. We emphasize again that the large values of $M$---required to resolve sharp features satisfactorily---are out of reach for previous methods due to their inherent $O(DM^2)$ scaling \cite{KITE_2020,Garcia_PRL/2015, FAN20211}. \textit{FastCheb} offers the possibility of evaluating $\sigma_{xy}(E)$ with high precision and energy resolution, as demonstrated here, being an order-$D M \log M$ approach. In fact, its full-spectral capability is essential to capture the intrinsic Hall conductivity due to being a bulk Fermi sea property. This contrasts with the case of the longitudinal conductivity (a bulk Fermi surface property) where studies for a small grid of Fermi energies can be carried out efficiently  with the single-shot algorithm   \cite{Ferreira/2015}. Here, the unique advantage of \textit{FastCheb} lies in the ability to efficiently reconstruct $\sigma_{aa}(E)$ over the \textit{full} spectrum. In this example, a speed-up of up to 10 over the standard method can be achieved \cite{supp_material}.


\textit{Final remarks.---} Applying the fast Fourier-Chebyshev algorithm to the study of quantum-critical phenomena can be a starting point for future research. Specifically, the investigation of the scaling behavior of Anderson transitions (such as metal-to-insulator and integer quantum Hall transitions \cite{QH_Evers/2008}) could benefit enormously from the spectral machinery developed in this work. In fact, the size of the lattice in our 2D conductance study [see Figs. \ref{fig:01}(b)-(d)] is already larger than recent calculations for semiclassical network models \cite{QH_Dresselhaus/2022}, which is promising. \textit{FastCheb} could also facilitate finite-size scaling analysis of the Hall conductivity, far less studied numerically than the diagonal response but equally important.


\textit{Acknowledgements.---} SGdC and DAB acknowledge support from CAPES (88887.510399$/$2020-00), CAPES-PRINT (88887.310281$/$2018-00),  CNPq (309835$/$2021-6) and Mackpesquisa. Supercomputer time was provided on MackCloud and Viking (University of York). JMVPL acknowledges the Centro de Física do Porto funded by the Portuguese Foundation for Science and Technology (FCT) within the Strategic Funding UIDB$/$04650$/$2020. A.F. acknowledges support from the Royal Society (URF$\backslash$R$\backslash$191021 and RF$\backslash$ERE$\backslash$210281). SGdC is grateful to Jesuel Marques Leal Junior and Carlos Guillermo Giménez de Castro for useful discussions.

\bibliography{BibMaterials}

\clearpage
\onecolumngrid

\section*{Supplementary Material}

\subsection*{Kubo-Bastin framework}  

The DC conductivity response tensor at finite temperature in the Kubo-Bastin formulation may be cast as 
\begin{align}
\sigma_{ab}(\mu,T)=\frac{2e^{2}\hbar}{\pi\Omega}\int dE\,f(E,\mu,T)\,\Re\,\textrm{Tr}\left[i\hat{v}_{a}\hspace{1mm}\frac{d\hat{\mathcal{G}}(E)}{dE}\hspace{1mm}\hat{v}_{b}\,\Im\,\hat{\mathcal{G}}(E)\right].
\end{align}

An $O(DM^2)$-implementation of the Kubo-Bastin formula was presented in Ref. \cite{Garcia_PRL_supp/2015}. We start, as in Ref. \cite{Garcia_PRL_supp/2015}, by approximating the  Green's functions as a truncated Chebyshev series via the kernel polynomial method. Performing the double-Chebyshev expansion and exploiting the stochastic trace technique, we arrive at
\begin{equation}
    \sigma_{a b}(\mu,T)\simeq i\frac{8e^2}{h \Omega}\int^1_{-1}d\varepsilon f(\varepsilon, \tilde \mu, \tilde T) \braket{\braket{\hspace{1mm}\tilde{\zeta}^r_{a b}(\varepsilon)\hspace{0.8mm}}}_R,
\label{eq12:BastinExpanded}
\end{equation}
where $\tilde T$ and $\tilde \mu$ are the dimensionless temperature and chemical potential variables and $\tilde{\zeta}^r_{a b}(\varepsilon)$ is given by:
\begin{equation}
\tilde{\zeta}^r_{a b}(\varepsilon)=\sum^{M-1}_{m,n=0}g_{mn}(\varepsilon)\mu^{r,a b}_{mn},\,\,\,\,\,\,\,\,\,\,\,\,\mu^{r,a b}_{mn}=\bra{r} T_m(\hat H)\tilde{v}_a T_n(\hat H)\tilde{v}_b\ket{r}.
\label{eq13:Zeta}
\end{equation}

The energy dependent pre-factors read as
\begin{align}
    g_{mn}(\varepsilon)\equiv  g_m g_n (1-\varepsilon^2)^{-2} \times
    [\hspace{0.8mm}&(\varepsilon-in\sqrt{1-\varepsilon^2})e^{in\,\arccos (\varepsilon)}T_m(\varepsilon)+ \label{eq14:Bastinamma}  \\
     &(\varepsilon+im \sqrt{1-\varepsilon^2})e^{-im\,\arccos(\varepsilon)}T_n(\varepsilon) \hspace{0.8mm}]. \nonumber
\end{align}

Next, we cast Eq. (\ref{eq13:Zeta}) in terms of scalar products of energy-space vectors as done for the DC longitudinal conductivity in the main text. This time, however, 6 energy-space vectors are needed as opposed to only 2 for the Kubo-Greenwood formula. By choosing $\varepsilon_k=\cos{(\pi (k+1/2)/M)}$, the exponential terms reduce to $e^{\mp im\,\arccos(\varepsilon_k)}= e^{\mp i \frac{\pi m(k+1/2)}{M}}$. Expanding out $g_{mn}(\varepsilon)$ and $\mu^{r, a b}_{mn}$ in Eq. (\ref{eq13:Zeta}) followed by some simplifications yields:

\begin{nalign}
\tilde{\zeta}^r_{a b}(\varepsilon_k) = \varepsilon_k &\left[\braket{\phi^{L,3;r}_{a;k} | \phi^{R,1;r}_{b;k}} + \braket{\phi^{L,1;r}_{a;k}| \phi^{R,3;r}_{b;k}}\right] +\\
i\sqrt{1-\varepsilon^2_k}&\left[ \braket{\phi^{L,2;r}_{a;k} | \phi^{R,3;r}_{b;k}} -\braket{\phi^{L,3;r}_{a;k} | \phi^{R,2;r}_{b;k}}\right], \label{eq15:EspaceZeta}
\end{nalign}
with the following energy-space vectors ($\chi=a,b$):
\begin{align}
\ket{\phi^{R/L,1;r}_{\chi;k}}&=\sum^{M-1}_{m=0}e^{i\frac{\pi mk}{M}}c_m g_m\ket{\chi^{R/L;r}_m}, \nonumber \\
\ket{\phi^{R/L,2;r}_{\chi;k}}&=\sum^{M-1}_{m=0}e^{i\frac{\pi mk}{M}}c_m g_m m\ket{\chi^{R/L;r}_m}, \label{eq16:BastinEvecs}\\
\ket{\phi^{R/L,3;r}_{\chi;k}}&=\sum^{M-1}_{n=0}\frac{\cos\left(m\pi (k+1/2)/M\right)}{1+\delta_{m,0}} \frac{2K_m}{\pi}\ket{\chi^{R/L;r}_m}. \nonumber
\end{align}

The auxiliary vectors are  $|a_{m}^{L;r}\rangle=T_{m}(\hat H)|r\rangle$ and $|a_{n}^{R;r}\rangle=\tilde{v}_{a}T_{n}(\hat H)\tilde{v}_{b}|r\rangle$. Note that these extra energy-space vectors do not affect the favourable $O(D M \log M)$-scaling of \textit{FastCheb}. One may traverse the $\boldsymbol{a}_R$ and $\boldsymbol{a}_L$ matrices performing the FFTs and immediately updating the partial results of the scalar products from Eq. (\ref{eq15:EspaceZeta}):
\begin{nalign}
p_i(\varepsilon_k) &= p_{i-1}(\varepsilon_k)+ \phi^{L,3;r\dagger}_{a;i,k} \, \phi^{R,1;r}_{a;i,k}+ \phi^{L,1;r\dagger}_{a;i,k}  \phi^{R,3;r}_{a;i,k}, \\
 w_i(\varepsilon_k) &= w_{i-1}(\varepsilon_k)+\phi^{L,2;r\dagger}_{a;i,k} \phi^{R,3;r}_{a;i,k}-\phi^{L,3;r\dagger}_{a;i,k}  \phi^{R,2;r}_{a;i,k}.
 \label{eq17:BastinRowOps}
\end{nalign}

Once the row-wise sweep is completed, the contribution of the random vector to the response function is
\begin{equation}
\tilde{\zeta}^r_{\alpha \beta}(\varepsilon_k)=\varepsilon_k p_D(\varepsilon_k)+i\sqrt{1-\varepsilon_k^2}\,w_D(\varepsilon_k).
\label{eq17:finalZeta}
\end{equation}
The Kubo-Bastin FFT processing of the matrices $\boldsymbol{a}_{R(L)}$ takes up to 3 times longer than the Kubo-Greenwood case. However, the scaling of the CPU time remains bounded by $O(D M \log M)$.

\begin{figure}
\begin{center}
\resizebox{15cm}{!} {\includegraphics{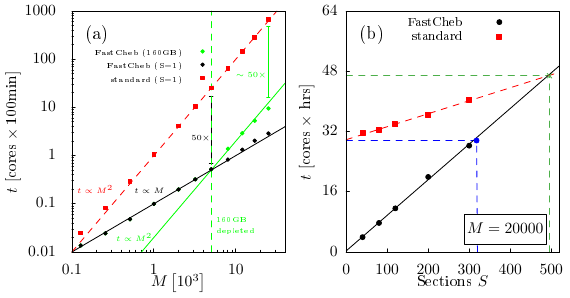}}
\caption{\textbf{(a)} Extrapolation of runtime $t$ for a graphene system with $D=10^6$ from the scaling law in Fig 1.(a) (main text). For the performance comparison, a hypothetical memory cutoff of 160 GB is considered. Thus, the maximum  number of Chebyshev vectors stored on RAM is $M_{\text{cut}}=5000$. For $M>M_{\text{cut}}$, partitioning is required. \textit{FastCheb} is predicted to exhibit an excellent performance in all cases, providing a speed-up of about 50 for large $M$ values.  \textbf{(b)} Evolution of $t$ with number of memory partition sections $S$ for transport simulations run with $D=10^4$ and $M=20000$.}
\label{fig:03}  
      \end{center}  
 \end{figure}

\subsection*{Memory management: the partitioned fast-Fourier Chebyshev method}  
 
For large problems it becomes impractical to store the entire matrices $\boldsymbol{a}_{R(L)}$. Some sort of partitioning needs to be implemented to mitigate the memory cost. One may divide $\boldsymbol{a}_{R(L)}$ into $S$ blocks  each containing $I=D/S$ rows. Series of vector recursions are then repeated $S$ times, storing in RAM the $s$-th block from each matrix at every iteration. The FFT part is performed on those blocks alone, updating $p_{sI}(\varepsilon_k)$ to $p_{(s+1)I}(\varepsilon_k)$. The memory cost is thus reduced to $O(2DM/S)$, at the expense of a higher algorithmic complexity. The CPU time now scales as $O(2SZDM)$, where $Z$ is the average coordination number of the lattice.

The CPU scaling laws obtained in our benchmark [Fig. \ref{fig:01}(a) (main text)] can be used to extrapolate the run times for larger simulations.  For illustration purposes, we consider a system with $D=10^6$ orbitals and assume that a maximum of 160 GB of RAM is available. In this scenario, memory becomes depleted when $M=M_{\text{cut}}=5000$ (this assumes double complex precision). For $M>M_{\text{max}}$, partitioning is used in \textit{FastCheb} to keep memory usage below the threshold of 160 GB. Figure \ref{fig:03}(a) shows the predicted run times in \textit{FastCheb}, partitioned \textit{FastCheb} ($S>1$) and the standard spectral method. Due to the memory restriction, the number of blocks $S$ is progressively increased when $M$ increases: $S=2$ for $M=8000$, $S=3$ for $M=12000$, $S=4$ for $M=17000$ and $S=5$ for $M=25000$. Because $S \propto M$, the computational complexity in the partitioned \textit{FastCheb} now scales as $O(2SDM)= O(DM^2/I)$, with $I \equiv M/S > 1$. Thus, a massive gain over the standard approach is  obtained so long a large number of $D$-dimensional Chebyshev vectors can be stored in memory (i.e. $I\gg 1$). 
 
In the standard method, a similar partitioning technique may be employed. Column blocks are stored instead, and only one vector recursion needs to be repeated per block. The complexity scales as $O(SZDM+DM^2)$. In Fig. \ref{fig:03}(b), we compare this approach against the partitioned \textit{FastCheb}. Despite repeating a vector recursion an extra time per section, \textit{FastCheb} retains most of its advantage, even for large values of $S$. When $S=40$, \textit{FastCheb} is over 10 times faster than the partitioned standard. The blue dot marks the $S$-threshold at which the FFT takes the same time as the standard method with $S=1$. This happens for  $S=312$, meaning that \textit{ FastCheb could use 1/312 of the memory in the standard approach while requiring the same time to complete the calculation.} Eventually, for a heavily partitioned simulation, the cost of performing an extra recursion for each partition in the FFT algorithm surpasses the baseline cost of the matrix product cost in the standard method. That threshold is indicated by the green point.

\newpage 

\subsection*{Additional details}

Tables \ref{tab1} and \ref{tab2} summarize the technical data and parameters in the simulations presented in the main text. Cost in \ref{tab2} refers to the approximate total CPU time for a single random vector realization, RAM indicates the peak memory usage, and ''Est. speedup'' indicates the estimated speedup with respect to the standard kernel polynomial method.
\vspace{10 cm}

\begin{table}
\caption{List of CPUs utilized in this work.}
\noindent
\begin{tabular}{|l||*{1}{c|}}\hline
&\makebox[3em]{Specifications}\\\hline\hline
CPU 1 & Intel(R) Xeon(R) CPU E5-2630 v4 @ 2.20GHz\\\hline
CPU 2 & Intel(R) Xeon(R) 6138 20-core@2.0 GHz\\\hline
CPU 3 & Intel(R) Xeon(R) Platinum 8160 4x24-core@2.1GHz \\\hline
\end{tabular}
\label{tab1}
\end{table}

\begin{table}
\caption{\textit{FastCheb} speed-up and RAM usage}
\noindent
\begin{tabular}{|l||*{5}{c|}}\hline
Simulation&\makebox[3em]{CPU}&\makebox[3em]{R}&\makebox[3em]{S}&\makebox[3em] {Cost/RAM}&\makebox[6em] {Est. Speedup}\\\hline\hline
Fig. \ref{fig:01}(b), M=10000  & CPU 1 & 40 & 10 & 41min/-& 1.5$\times$\\\hline
Fig. \ref{fig:01}(b), M=20000  & CPU 1 & 40 & 10 & 84min/-& 6 $\times$\\\hline
Fig. \ref{fig:01}(b), M=40000  & CPU 1 & 40 & 10 & 290min/-& 9$\times$\\\hline
Fig. \ref{fig:01}(b), M=100000 & CPU 3 & 40 & 4  & 75min, 750GB& 100$\times$\\\hline\hline

Figs. \ref{fig:01}(c)-(d), M=10000 & CPU 2 & 180 & 1 & 15min/-& 10$\times$\\\hline
Figs. \ref{fig:01}(c)-(d), M=15000 & CPU 2 & 180 & 2 & 42min, 90GB& 3$\times$\\\hline
Figs. \ref{fig:01}(c)-(d), M=28000 & CPU 2 & 180 & 1 & 42min, 330GB& 10$\times$\\\hline
Figs. \ref{fig:01}(c)-(d), M=56000 & CPU 2 & 106 & 1 & 90min, 700GB& 20$\times$\\\hline\hline

Fig. \ref{fig:02}, M=5000 & CPU 2 & 40 & 5 & 30min, 306GB& 8-10$\times$\\\hline
Fig. \ref{fig:02}, M=10000 & CPU 2 & 40 & 10 & 100min, 306GB& 8-10$\times$\\\hline
Fig. \ref{fig:02}, M=20000 & CPU 2 & 40 & 20 & 5h, 306GB& 8-10$\times$\\\hline
Fig. \ref{fig:02}, M=40000 & CPU 2 & 40 & 40 & 20h, 306GB& 8-10$\times$\\\hline
\end{tabular}
\label{tab2}
\end{table}

\end{document}